# Lessons from Hubble and Spitzer: 1D Self-consistent Model Grids for 19 Hot Jupiter Emission Spectra

Lindsey S. Wiser[1], Michael R. Line[1], Luis Welbanks[1,6], Megan Mansfield[2,6], Vivien Parmentier[3], Jacob L. Bean[4], and Jonathan J. Fortney[5]
[1] School of Earth and Space Exploration, Arizona State University, 781 Terrace Mall, Tempe, AZ 85287, USA; lindsey.wiser@asu.edu
[2] Steward Observatory, University of Arizona, Tucson, AZ 85719, USA
[3] Université Côte d'Azur, Observatoire de la Côte d'Azur, CNRS, Laboratoire Lagrange, France
[4] Department of Astronomy and Astrophysics, University of Chicago, 5640 S Ellis Ave, Chicago, IL 60637, USA
[5] Department of Astronomy and Astrophysics, University of California, Santa Cruz, CA 95064, USA
Received 2023 September 18; revised 2024 April 26; accepted 2024 May 1; published 2024 August 2

## Abstract

We present a population-level analysis of the dayside thermal emission spectra of 19 planets observed with Hubble WFC3 and Spitzer IRAC 3.6 and 4.5 $\mu$m, spanning equilibrium temperatures 1200–2700 K and 0.7–10.5 Jupiter masses. We use grids of planet-specific 1D, cloud-free, radiative–convective–thermochemical equilibrium models (1D-RCTE) combined with a Bayesian inference framework to estimate atmospheric metallicity, the carbon-to-oxygen ratio, and day-to-night heat redistribution. In general, we find that the secondary eclipse data cannot reject the physics encapsulated within the 1D-RCTE assumption parameterized with these three variables. We find a large degree of scatter in atmospheric metallicities, with no apparent trend, and carbon-to-oxygen ratios that are mainly consistent with solar or subsolar values but do not exhibit population agreement. Together, these indicate either (1) formation pathways vary over the hot and ultra-hot Jupiter population and/or (2) more accurate composition measurements are needed to identify trends. We also find a broad scatter in derived dayside temperatures that do not demonstrate a trend with equilibrium temperature. Like with composition estimates, this suggests either significant variability in climate drivers over the population and/or more precise dayside temperature measurements are needed to identify a trend. We anticipate that 1D-RCTE models will continue to provide valuable insights into the nature of exoplanet atmospheres in the era of JWST.

*Unified Astronomy Thesaurus concepts:* Exoplanet atmospheres (487); Exoplanets (498); Extrasolar gaseous giant planets (509); Bayesian statistics (1900); Exoplanet atmospheric composition (2021)

## 1. Introduction

Of the over 5000 confirmed exoplanet discoveries, hot Jupiters are among the most well-characterized (Fortney et al. 2021). Hot Jupiters are close-in gas giants with orbital periods of less than 10 days (equilibrium temperatures, $T_{\rm eq}$, >1000 K), making them ideal targets for repeated atmosphere observations with transit spectroscopy—both emission from the dayside and transmission of stellar light through the planet's limbs. Transit spectroscopy with the Hubble Space Telescope (HST) and the Spitzer Space Telescope has provided the observational foundation for understanding exoplanet atmosphere climates, compositions, and processes.

Key questions guiding our work include: What are the intrinsic elemental compositions (metallicity, C/O) of planetary atmospheres? Do these compositions provide insights into planet formation and migration mechanisms? How is stellar energy absorbed and transported by atmospheres? What mechanisms control the efficiency of energy transport?

### 1.1. Atmosphere Composition and Planet Formation

A planet's formation history is encoded in its atmosphere's composition (e.g., Stevenson & Lunine 1988; Mordasini et al. 2016). The core accretion model is the leading explanation for the formation of planets in our Solar System (Stevenson & Lunine 1988). In this model, planet formation starts in a protoplanetary disk at orbital radii where predominant molecular species condense into solids (i.e., "ice lines"), prompting dust and ice particles to come together to form the cores of future planets (Pollack et al. 1996). Atmosphere envelopes then accrete around these solid cores; therefore, the composition of a planet's atmosphere can be expected to reflect the composition of gas and polluting pebbles in the protoplanetary disk at the radius of formation (D'Angelo & Lubow 2008; Öberg et al. 2011). Following the core accretion model, it is thought that hot Jupiter formation may start with superstellar metallicity (M/H) cores beyond the H$_2$O ice line, followed by the accretion of substellar M/H gas to build up their expansive atmospheres. Migration inward can then facilitate M/H enhancement via the accretion of leftover disk material (e.g., Fortney et al. 2021). Over the Jovian-to-Neptune mass range, these processes are expected to result in solar to >100× solar M/H (Thorngren & Fortney 2019). Additionally, M/H is predicted to decrease with increasing planet mass due to the proportional impact of M/H enhancement during migration, i.e., more massive gas giants have extended H/He atmospheres that dilute the accretion of metals. Decreasing M/H with planet mass is a trend in our Solar System (e.g., Kreidberg et al. 2014; Guillot & Gautier 2015) and has been extrapolated to apply to hot Jupiters (e.g., Fortney et al. 2013; Kreidberg et al. 2014; Mordasini et al. 2016; Thorngren et al. 2016; Thorngren & Fortney 2019; Welbanks et al. 2019).

The atmospheric carbon-to-oxygen ratio (C/O) is a complementary metric hypothesized to be sensitive to the radial

---

location within a disk at which a planet accreted its volatiles (Madhusudhan 2012). From Öberg et al. (2011), beyond the $H_2O$ ice line, atmospheres form from supersolar C/O gas and subsolar C/O grains, with the collection of additional gas and pebbles facilitated by migration inward. This process should result in an atmospheric C/O roughly consistent with stellar ratios. Significantly higher C/O ($\gtrsim 1.0$) may require the accretion of primarily C-rich gas beyond the $H_2O$ ice line or at even farther orbital distances beyond the $CO_2$ ice line (Booth et al. 2017). Supersolar C/O can also be assisted by significant amounts of C-rich grain accretion internal to the $H_2O$ ice line during migration inward in an undissipated disk. Low C/O ($\lesssim 0.1$) necessitates envelopes dominated by O-rich icy planetesimal impacts (Mordasini et al. 2016).

### 1.2. Heat Transport and Climate

The efficiency of heat transport is an indicator of climate processes, including planetary winds and aerosols (clouds and hazes), which are pervasive in exoplanet atmospheres (e.g., Iyer et al. 2016; Sing et al. 2016). Three-dimensional general circulation models (GCMs) of tidally locked hot Jupiters predict a trend of increasing day–night temperature contrasts with increasing planet equilibrium temperature ($T_{eq}$). This is due to increased dayside irradiation and the < 1 ratio of radiative cooling timescales to the timescales of advective (wind) and gravity wave transport circulating heat from day to night (e.g., Showman & Guillot 2002; Perez-Becker & Showman 2013; Komacek & Showman 2016). In other words, the increase in radiative cooling with $T_{eq}$ is greater than the balancing effect of heat circulation. However, cloudless GCMs systematically underpredict this day–night temperature contrast when compared to exoplanet phase curve observations that plot a planet's variation in brightness temperature with longitude (e.g., Kataria et al. 2015; Lewis et al. 2017; Parmentier & Crossfield 2018; Zhang et al. 2018). The presence of nightside clouds could explain this phenomenon—Parmentier et al. (2021) use cloud-free and cloudy GCMs to show that, in the absence of nightside clouds, heat redistribution is efficient at $T_{eq} < 1600$ K, with decreasing efficiency up to $T_{eq} = 2200$ K. Nightside clouds, however, restrict heat redistribution at all equilibrium temperatures. Clouds may induce a warming greenhouse effect beneath them but restrict nightside radiative cooling. Heat is, therefore, circulated back to the dayside, which continues to warm relative to the planet's equilibrium temperature, resulting in large day–night temperature contrasts, i.e., poor day-to-night heat transport.

Aerosols remain one of the largest outstanding uncertainties in interpreting exoplanet spectra, yet they significantly impact temperature structure and atmosphere dynamics (e.g., Gao et al. 2021). Complicating the heat transport picture, a diversity of cloud/aerosol molecular compositions and particle sizes may exist throughout the hot and ultra-hot Jupiter population (Lodders & Fegley 2006; Marley & Robinson 2015). Furthermore, cooling from the dissociation of molecular species such as $H_2$ on a planet's dayside, accompanied by heating from molecular recombination on the nightside, might reduce temperature contrasts and increase day-to-night heat transport (e.g., Tan & Komacek 2019; Mansfield et al. 2020; Roth et al. 2021) relative to the decrease in heat transport resulting from nightside clouds.

An atmosphere's vertical temperature structure is also a key measurable property because it is diagnostic of the relative balance between absorbed stellar energy and reradiated planetary thermal energy. Temperature inversions in the upper atmospheres of planets hotter than ∼2000 K have been predicted and supported by observations thus far (e.g., Hubeny et al. 2003; Fortney et al. 2008; Arcangeli et al. 2018; Lothringer et al. 2018; Mansfield et al. 2022; Coulombe et al. 2023). These inversions may be explained by the presence of significant absorbers such as TiO, VO, SiO, atomic metals, metal hydrides, and continuous opacities, which condense out in cooler atmospheres.

### 1.3. Modeling Approach

A broad range of modeling approaches have been used to constrain the properties of exoplanet atmospheres (e.g., Madhusudhan 2018). Typically, methods either fall into the forward modeling category, which aims to predict atmosphere properties via comparisons of precomputed models to observations, or "retrievals," which typically make fewer model assumptions and focus on statistical fitting to data. Forward models range from 1D, changing only with altitude along a pressure–temperature profile, to computationally intensive 3D GCMs that include latitudinal and longitudinal effects such as equatorial jets, condensation along the day–night terminator, and more (Madhusudhan 2019). Forward models are typically self-consistent, meaning they follow a set of interdependent conditions, in our case, radiative–convective–thermochemical equilibrium (RCTE). Retrievals also vary in complexity (e.g., chemical equilibrium versus "free" chemistry, 1D versus multidimensional), but generally, they add more free parameters at the expense of possible unphysical solutions.[7]

Historically, 1D-RCTE models have been used to predict and interpret data for the atmospheres and climates of Earth (e.g., Manabe & Wetherald 1967; Hansen et al. 1981), Solar System bodies (e.g., Appleby 1986; McKay et al. 1989; Marley & McKay 1999; von Paris et al. 2015), stars (e.g., Kurucz 1979; Husser et al. 2013), brown dwarfs (e.g., Allard et al. 2003; Marley & Robinson 2015), hot Jupiter exoplanets (e.g., Burrows et al. 2000; Fortney et al. 2008), terrestrial exoplanet habitability (e.g., Kasting et al. 1993; Rugheimer & Kaltenegger 2018), and the thermal evolution of substellar objects (e.g., Fortney et al. 2007; Marley et al. 2021). However, retrievals have increasingly become commonplace in exoplanet studies. Retrievals are computationally efficient, and there continue to be many uncertainties in our understanding of exoplanet atmosphere physics; retrievals may illuminate explanations for observed spectral features that more restrictive methods exclude (e.g., Sheppard et al. 2017; MacDonald & Madhusudhan 2019; Chubb et al. 2020; Sotzen et al. 2020). Unfortunately, retrieval results may be nonphysical, and many free parameters can lead to overfitting. Recently, Bayesian statistical methods combined with 1D-RCTE forward models have started to be applied to exoplanet observations (e.g., Arcangeli et al. 2018; Mansfield et al. 2018; Glidic et al. 2022; Bell et al. 2023; Coulombe et al. 2023). This method relies on self-consistent physics to avoid nonphysical models, and it is computationally efficient due to utilizing precomputed 1D models combined with Bayesian statistical methods. Ultimately, utilizing various data–model inference approaches is critical for comparing model results and developing a complete picture of exoplanet atmospheres.

---

[7] A catalog of existing exoplanet atmospheric retrieval codes can be found in MacDonald & Batalha (2023).

Table 1
Data Sources for Each Planet in Our Population

| Planet | Hubble WFC3 | Spitzer 3.6 and 4.5 $\mu$m |
| --- | --- | --- |
| HD189733b | Crouzet et al. (2014) | 3.6: Knutson et al. (2012), 4.5: Bell et al. (2021) |
| WASP-43b | Kreidberg et al. (2014) | Stevenson et al. (2017) |
| HD209458b | Line et al. (2016) | Diamond-Lowe et al. (2014) |
| CoRoT-2b | Wilkins et al. (2014) | Deming et al. (2011) |
| TrES-3b | Ranjan et al. (2014) | Fressin et al. (2010) |
| WASP-77b | Mansfield et al. (2022) | Mansfield et al. (2022) |
| WASP-4b | Ranjan et al. (2014) | Beerer et al. (2011) |
| WASP-79b | Mansfield et al. (2021) | Garhart et al. (2020) |
| HAT-P-32b | Nikolov et al. (2018) | Zhao et al. (2014) |
| WASP-74b | Mansfield et al. (2021) | Fu et al. (2021) |
| HAT-P-41b | Mansfield et al. (2021) | Garhart et al. (2020) |
| KELT-7b | Mansfield et al. (2021) | Garhart et al. (2020) |
| WASP-76b | Mansfield et al. (2021) | May et al. (2021) |
| HAT-P-7b | Mansfield et al. (2018) | Wong et al. (2016) |
| WASP-121b | Mansfield et al. (2021) | Garhart et al. (2020) |
| WASP-18b | Arcangeli et al. (2018) | Maxted et al. (2013a) |
| WASP-103b | Kreidberg et al. (2018) | Kreidberg et al. (2018) |
| WASP-12b | Stevenson et al. (2014) | Stevenson et al. (2014) |
| WASP-33b | Haynes et al. (2015) | Zhang et al. (2018) |

**Notes.** Each HST data set is presented in Mansfield et al. (2021), with the addition of WASP-77b from Mansfield et al. (2022). Spitzer data come from assorted sources.

### 1.4. In This Work

The goals of this investigation are: (1) to determine if the 1D-RCTE assumption is valid for the daysides of hot Jupiters and whether thermal emission spectra arising from said assumptions can provide adequate agreement to the observations, and (2) to self-consistently derive basic atmospheric properties, including metallicity, C/O, and dayside temperatures/heat transport. Many unanswered questions remain regarding the composition of exoplanet atmospheres, how composition connects to formation, and the mechanisms controlling heat transport. However, population analyses provide a foundation for future investigations.

We model dayside thermal emission spectra for 19 hot Jupiters with equilibrium temperatures 1200–2700 K and masses 0.7–10.5 $M_{\rm Jupiter}$. We apply a uniform modeling and parameter estimation approach to each planet, which we refer to as "grid-based retrieval." We produce grids of 1D-RCTE atmosphere models with varied elemental inventories (via metallicity and C/O) and dayside temperatures (via a heat redistribution parameter). All models are aerosol-free and specific to each planet's mass, radius, orbital parameters, and host star. We then use Bayesian statistical methods to identify and interpolate between models most consistent with spectral data. We present constraints on metallicity, C/O, and heat redistribution in the context of common population trend predictions from the literature. Finally, we discuss how this work complements other recent population studies, including those presented in Goyal et al. (2021) and Changeat et al. (2022).

### 2. Methods

#### 2.1. Data

We restrict our analysis to HST Wide Field Camera 3 (WFC3) observations using grism G141 1.1–1.7 $\mu$m (primarily sourced from Mansfield et al. 2021) and Spitzer InfraRed Array Camera (IRAC) 3.6 and 4.5 $\mu$m observations. Table 1 summarizes data references for each planet. WFC3's wavelength range is sensitive to water around 1.4 $\mu$m, while the IRAC wavelengths are sensitive to features of carbon-bearing species $CH_4$, CO, and $CO_2$, as well as temperature (Garhart et al. 2020). See Figure 1 for a visual of how spectra change with varied heat redistribution and composition parameters.

Generally, we take published values and uncertainties as they are and do not consider disparity over potential data artifacts, such as offsets between HST and Spitzer or under/overestimated error bars. However, two sources are referenced for HD189733b's Spitzer points—Bell et al. (2021) for 4.5 $\mu$m and Knutson et al. (2012) for 3.6 $\mu$m. Bell et al. (2021) perform a detailed comparison of observation-to-spectral point reduction methods at 4.5 $\mu$m and find a value within 1$\sigma$ agreement of the 4.5 $\mu$m result in Knutson et al. (2012). For WASP-76b, only a point at 4.5 $\mu$m is used because of significant disagreement in the literature over the value at 3.6 $\mu$m (Garhart et al. 2020; May et al. 2021). Finally, Kepler-13Ab, while included in Mansfield et al. (2021), is excluded from this manuscript because the data is visibly inconsistent with grid models; the Kepler-13Ab HST data are much hotter than our models predict.[8]

#### 2.2. Data–Model Inference Approach

We generate 1D-RCTE models using the ScCHIMERA framework (e.g., Arcangeli et al. 2018; Kreidberg et al. 2018; Mansfield et al. 2018; Piskorz et al. 2018; Baxter et al. 2020; Mansfield et al. 2021; Glidic et al. 2022; Mansfield et al. 2022; Coulombe et al. 2023), which was recently modified and benchmarked against M-dwarf systems (Iyer et al. 2023). Inputs of the 1D-RCTE code are elemental abundances, the top-of-atmosphere incident flux, and an internal temperature, which we set to 300 K. Varying the internal temperature alters the atmosphere's pressure–temperature profile at high pressures; however, we determined that those variations have a negligible impact on parameter estimations in this work because our grid-based retrievals to HST and Spitzer observations do not constrain internal temperature.[9] Elemental abundances are parameterized with the carbon-to-oxygen ratio (C/O) and metallicity ([M/H]), where the brackets denote $\log_{10}$ relative to solar and M is the sum of all non-H/He elements. All elemental abundances follow the Lodders et al. (2009) solar abundance pattern and are scaled together. For instance, if [M/H] = 1, all solar abundances for non-H/He elements are scaled by 10× (accounting for the renormalization of H). When adjusting C/O, the sum of C + O is preserved at the current metallicity scaling, and only the ratio between the two is adjusted. The incident stellar flux, given by the appropriate interpolated PHOENIX stellar model (Husser et al. 2013), is scaled first via the orbital semimajor axis and then by a heat redistribution parameter, $f$. In this paper, we use the heat distribution parameter from Arcangeli et al. (2018), defined as $(T_{day}/T_{eq})^4$. When $f = 1$, the dayside effective temperature is the equilibrium temperature, i.e., complete redistribution; the maximum theoretical value for $f$ is 2.66, denoting no heat redistribution (Cowan et al. 2012; Arcangeli et al. 2018). For each planet-specific model, these variables define the free parameter set: [M/H], C/O, and $f$. Planet system properties (semimajor axis, planet mass, radius, gravity, and stellar effective temperature) are taken from the Transiting Exoplanet Catalog (TEPCat;[10] Southworth 2011) and are summarized in Table 2.

---

[8] A brief discussion of Kepler-13Ab, including a figure of Kepler-13Ab data alongside grid models is presented in extended figures on the Zenodo repository: doi:10.5281/zenodo.11239517.

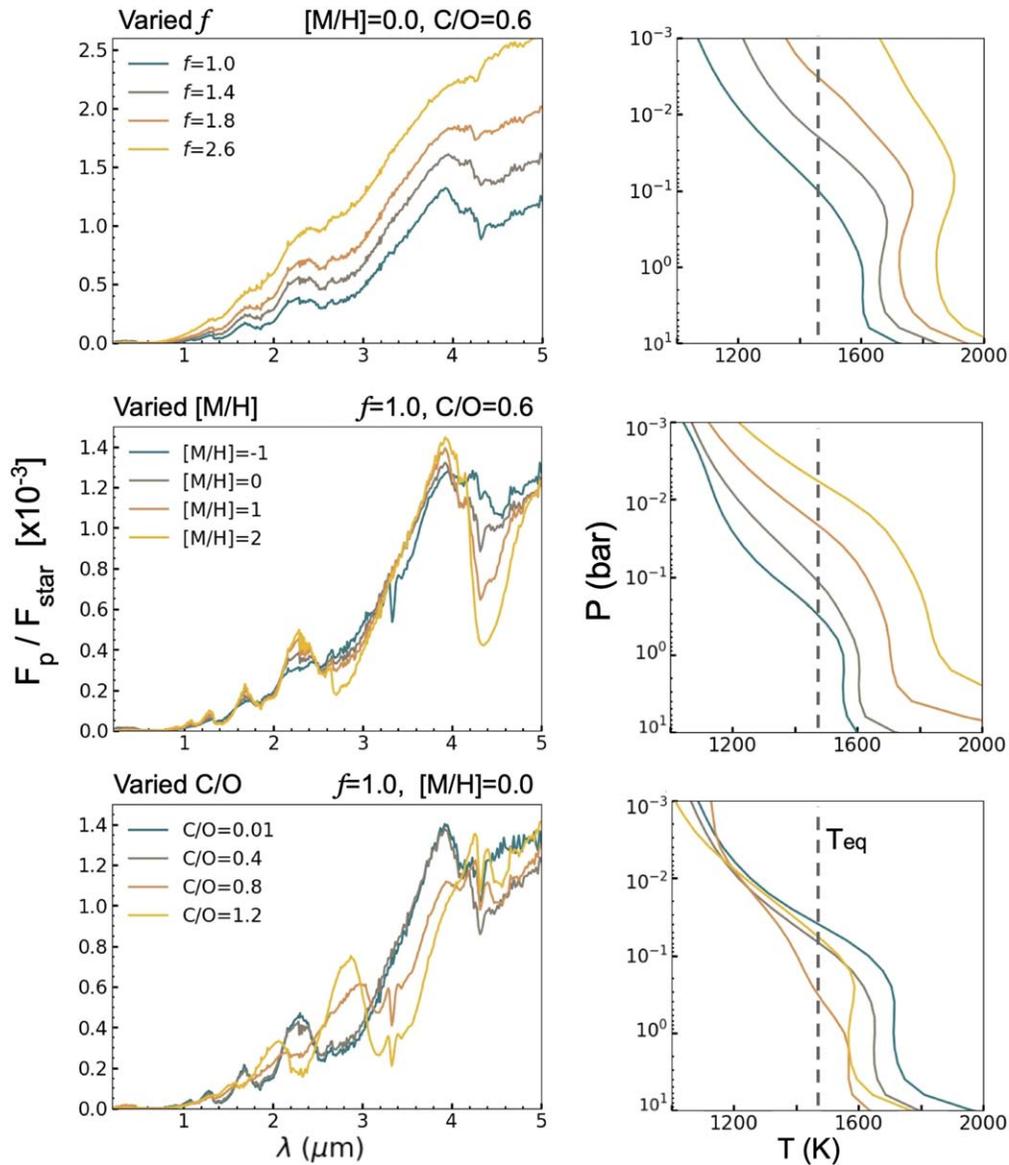

**Figure 1.** Select models from an HD209458b model grid ($T_{eq} = 1460$ K) to illustrate the impact of each grid parameter—dayside temperature via the heat redistribution parameter ($f$), $\log_{10}$ metallicity relative to solar ([M/H]), and C/O—on the modeled emission spectrum and pressure–temperature profiles.

For a given set of [M/H], C/O, and $f$, the ScCHIMERA 1D-RCTE routine iteratively solves for the radiative–convective–equilibrium pressure–temperature atmosphere structure and thermochemical equilibrium gas volume mixing ratios. Cloud-free radiative transfer is based upon the Toon et al. (1989) two-stream source function method, which calculates layer-by-layer radiative fluxes from a planet's internal thermal emission and an incident stellar flux, approximating a dayside hemispheric mean. Models are computed over the pressure range $10^{2.6}$ bar to $10^{-6.0}$ bar in layers $10^{0.2}$ bar thick. Given optical depths in each atmosphere layer, the equilibrium temperature structure can be solved using a Newton–Raphson iteration scheme (McKay et al. 1989). Optical depths at each layer are computed within the random-overlap resort-rebin correlated-K framework described in Amundsen et al. (2017) at spectral resolution $R = 250$, with a 10-point double Gaussian quadrature integration scheme for each k-distribution. Given the [M/H] and C/O-scaled elemental abundances, the gas mixing ratios used to compute these optical depths are derived from the NASA CEA2 routine, which seeks to minimize Gibbs-free energy in each atmosphere layer (Gordon & Mcbride 1994), combined with elemental rainout due to condensate formation. We include opacity sources from $H_2O$, $CO$, $CO_2$, $CH_4$, $NH_3$, $H_2S$, $PH_3$, HCN, $C_2H_2$, TiO, VO, SiO, FeH, CaH, MgH, CrH, AlH, Na, K, Fe, Mg, Ca, C, Si, Ti, O, $Fe^+$, $Mg^+$, $Ti^+$, $Ca^+$, $C^+$, $H_2$-$H_2$/He CIA, $H^-$ bound-free and free–free, $H_2$/He Rayleigh scattering, and additional UV opacities for CO, SiO, and $H_2$. Infrared absorption cross-section data are primarily sourced from the ExoMol project—see Mansfield et al. (2021) and Iyer et al. (2023) for details on individual line sources. Atomic and molecular UV line data are taken from the Kurucz website.[11]

---

[9] See extended figures on the Zenodo repository (doi:10.5281/zenodo.11239517) for tests comparing different internal temperatures.

[10] TEPCat: https://www.astro.keele.ac.uk/jkt/tepcat/

For each planet, we generate an individual 1D-RCTE model for each [M/H], C/O, and $f$ combination and the resultant top-of-atmosphere upwelling planetary flux. This process results in $\sim$2000–3000 models per planet (see Table 2). Finally, we include a parameter, $A$, which is a scale factor multiplying the planetary flux spectrum to account for temperature heterogeneities across the visible dayside disk (e.g., Taylor et al. 2020) and potential uncertainties in planet/star properties. To estimate planet parameters via data–model comparisons, we use the *PyMultiNest* routine for Bayesian nested sampling (Buchner et al. 2014) and trilinearly interpolate between modeled spectra using a modified version of the scipy.interpolate.griddata routine to minimize $\chi^2$ between model and data. Fisher & Heng (2022), based on an analysis of the Goyal et al. (2019) grid, advise caution using nested sampling and interpolation, in particular for parameters with nonlinear effects such as C/O and metallicity, showing that retrieved parameters may have offsets from the true value. To mitigate possible offsets, we employ finer grid spacing than in Goyal et al. (2019), as well as fewer varied grid parameters. We tested the accuracy of the trilinear interpolation and found that the differences between an exact model and the interpolated model were negligible (relative to the reported uncertainties on the data), and hence do not bias our inferences.

We consider a parameter to be constrained if the probability at the edge of the prior for that parameter is 70% lower than the probability at the median of the posterior probability distribution. In some cases, this returns a partial constraint with an upper or lower limit. In the case of a bimodal distribution, we present both possible solutions by splitting the *PyMultiNest* samples at the point of lowest probability between solutions and compute each solution's probability distribution independently.

To assess the dependency of parameter constraints on various model assumptions and on the data, we explore the following scenarios:

1. *Fiducial*. Estimate [M/H], C/O, $f$, and $A$. In this manuscript, each figure presents only the fiducial scenario, with the exception of Figure 10.
2. *Fixed to solar composition*. Fix [M/H] = 0.0 and C/O = 0.55, i.e., solar composition, and estimate $f$ and $A$.
3. *Fixed $f$*. Fix heat redistribution to its predicted value for the presence of nightside clouds from Parmentier et al. (2021) and estimate $A$, [M/H], and C/O.
4. *Fixed scale (A)*. Fix $A = 1$ and estimate $f$, [M/H], and C/O.
5. *HST WFC3 only*. Estimate all four parameters, but exclude Spitzer IRAC data in order to test the dependence of parameter estimations on Spitzer. Combining observations from different instruments can introduce systematic incompatibilities (e.g., Yip et al. 2020); however, we note that initial JWST observations have been consistent with Spitzer (e.g., Ahrer et al. 2023; August et al. 2023; Bean et al. 2023; Coulombe et al. 2023; Kempton et al. 2023). Additionally, the Spitzer points in these data sets are most sensitive to possible 3D variations that we do not account for in our 1D models (e.g., Cowan et al. 2012).

### 3. Results

Here, we present results from our 1D-RCTE data–model comparisons. Figure 2 presents modeled spectra for each planet. In Section 3.1, we demonstrate that the 1D-RCTE assumption can adequately explain the observations. In Section 3.2, we discuss constraints on composition in the fiducial case, as parameterized by metallicity and C/O, and compare these constraints to population-level trends. Finally, in Section 3.3, we present and discuss constraints on the planetary climate-related parameters—heat redistribution and the scale factor.

#### 3.1. Quality of Data–Model Agreement

We quantify how well our 1D-RCTE grid-based retrievals agree with spectral observations using three common statistical tests: reduced-$\chi^2$ ($\chi^2_{\rm red}$), the Kolmogorov–Smirnov (K-S) statistic, and the Anderson–Darling (A-D) statistic. Reduced-$\chi^2$, or $\chi^2/N_{\rm DOF}$ where $N_{\rm DOF}$ is the number of degrees of freedom, is a common metric for justifying model fits. Unfortunately, $N_{\rm DOF}$ is nontrivial to define when our model (Line et al. 2013) is nonlinear (Andrae et al. 2010; Voinov et al. 2013). In this paper, we report $\chi^2_{\rm red}$ assuming $N_{\rm DOF}$ is the number of data points minus the number of varied grid parameters, but we additionally report the K-S and A-D tests for comparison. The K-S and A-D tests both compare histograms of normalized residuals ([data–model]/error) to a Normal distribution with mean = 0 and variance = 1 to quantify how likely those residuals are to be drawn from a Normal distribution, and therefore the likelihood that data variation about the model results from random measurement error (see Figure 3). Examining residuals makes the K-S test more reliable than $\chi^2_{\rm red}$, but the A-D test is more powerful still with added sensitivity to the tails of the distributions. For all three statistical tests, we compute $p$-values. The $p$-value represents the probability that a measurement with random error would be observed and produce that specific statistic value ($\chi^2_{\rm red}$, K-S, or A-D) if the best-fit model were true. For example, a $p$-value of 0.05 suggests that, over some number of repeated observations, only 5% of the time would the data–model combination produce that given statistic value. Therefore, $p$-values lower than a certain threshold are interpreted as a rejection of a null hypothesis. In model fitting, the null hypothesis is that the data are consistent with the model prediction. For the A-D test, a *p*-value can be approximated using a piece-wise function defined in D'Agostino & Stephens (1986).

---
[11] Kurucz website: http://kurucz.harvard.edu/.

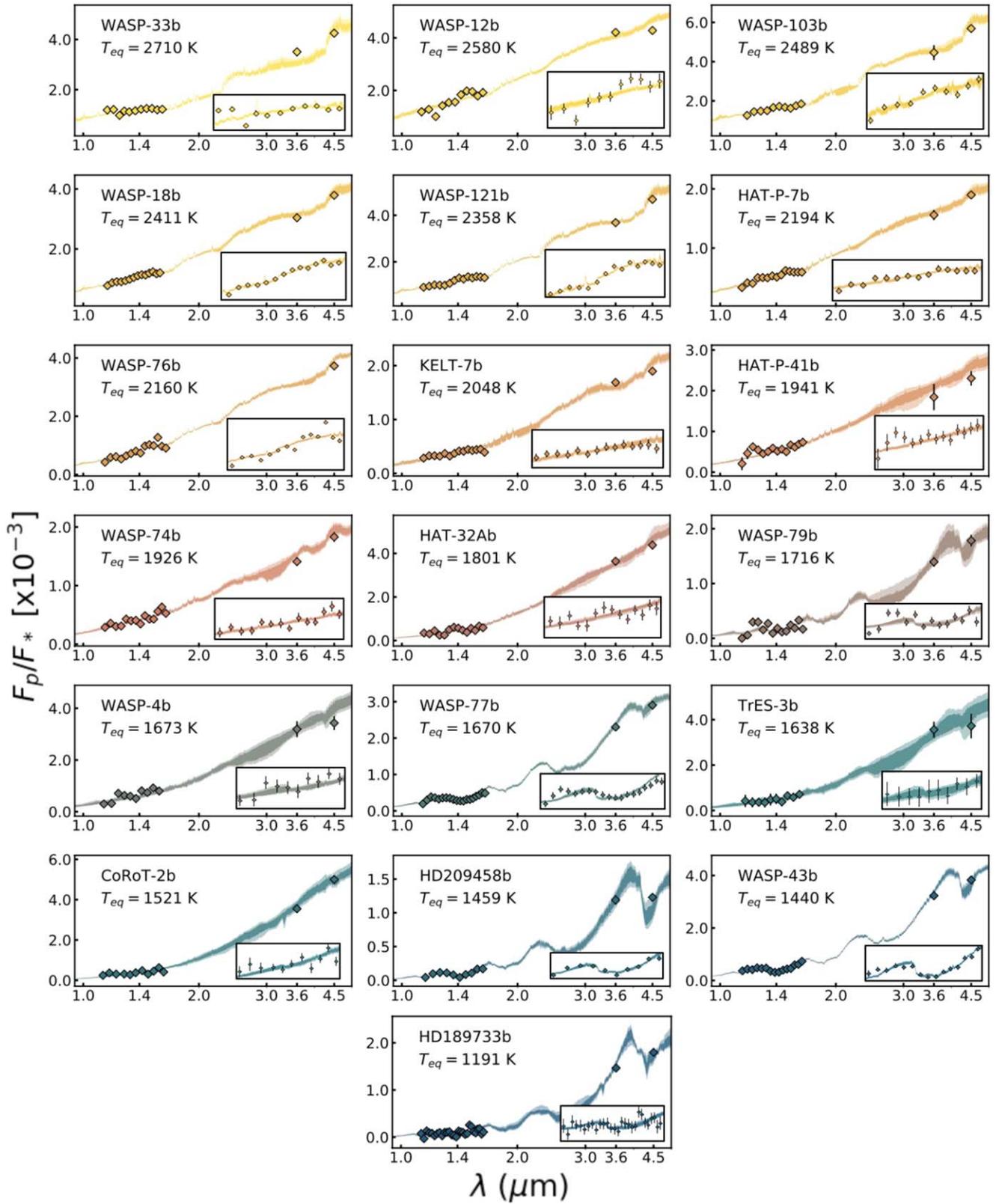

**Figure 2.** Modeled spectra for all 19 planets. The color bar correlates with planet equilibrium temperatures, with higher temperatures toward the top of the figure. Insets show HST WFC3 points, roughly 1.1–1.7 μm. This HST WFC3 wavelength range is more easily viewed in Figure 9. The 1σ and 2σ confidence regions calculated from Bayesian inference are shaded with darker and lighter shades, respectively. Corresponding corner plots for each planet can be found in the Zenodo repository (doi:10.5281/zenodo.11239517).

We take *p*-values > 0.05 to mean that the model assumptions cannot be rejected by the data. For all planets, the fiducial model assumptions are not rejected by the K-S and A-D tests, and only one data–model scenario is rejected by both tests (WASP-12b, $A = 1$ scenario). If one chooses to adopt 0.01 as a more restrictive *p*-value threshold (Wasserstein & Lazar 2016), no data–model scenarios for any planet are rejected by the A-D or K-S tests. This result suggests that the 1D-RCTE assumption produces models consistent with the HST and Spitzer observations. All four modeling assumptions we test (fiducial, fixed *f*, $A = 1$, and solar composition) cannot be ruled out with this data. In some cases, scenarios are in agreement—for example, if the composition is consistent with solar in the fiducial scenario, fixing [M/H] and C/O for the solar composition scenario will produce a model with similar parameters. Alternatively, some parameters are poorly constrained by the data (e.g., TrES-3b's composition is unconstrained over the full parameter range of its grid), allowing for close data–model agreement over a broad range of values. We also test the inclusion/exclusion of Spitzer observations. In some cases, excluding Spitzer does result in higher confidence in the model (larger *p*-value resulting from lower residuals); however, the loss of data points can also decrease confidence (lower *p*-value resulting from a smaller sample). The small number of data points considered in this study means we are unable to make statements regarding the reliability of Spitzer or any systematic incompatibilities from combining HST and Spitzer observations.

For most planets, all five data–model scenarios cannot be rejected (assuming the 0.05 *p*-value threshold) by at least two of the three statistical tests. Exceptions include WASP-12b, for which the $A = 1$ scenario is rejected by all three tests, and WASP-77b, for which the fixed *f* scenario is rejected by two out of three tests. There are three planets for which all *p*-values are higher than 0.05, i.e., no data–model scenarios can be rejected by any of the statistical tests: Kelt-7b, HAT-32Ab, and TrES-3b. All three planets have unconstrained or only partially constrained compositions in all scenarios. We refer the reader to Table 3 and Figure 10 in the Appendix for a closer look at statistical tests for all planets and model scenarios.

### 3.2. Composition

Based on *p*-value calculations (see Appendix Table 3 and Figure 10), solar composition cannot be ruled out for planets in this population. Therefore, population-level analysis relying on deviations from solar composition are inconclusive, but we can present C/O and [M/H] estimates from our grid-based retrievals to inform future observations and methods.

Both Solar System trends and the core accretion model applied to varied exoplanetary system architectures predict a downward trend in fractional metal content with increasing planet mass for gas giants (e.g., Fortney et al. 2013; Kreidberg et al. 2014; Guillot & Gautier 2015; Mordasini et al. 2016; Thorngren et al. 2016; Thorngren & Fortney 2019; Welbanks et al. 2019). In Figure 4, atmosphere metallicity estimations for this population are plotted alongside predicted trends from literature: exoplanet bulk metallicity predictions from a population study with thermal and structural evolution models (Thorngren et al. 2016), Solar System trends extended to higher planet masses (Kreidberg et al. 2014; Mordasini et al. 2016), and a trend in $H_2O$ abundance from a retrieval analysis of exoplanet transmission spectra (Welbanks et al. 2019). Welbanks et al. (2019) find that, while abundances for elements such as K and Na generally agree with Solar System predictions, $H_2O$ abundances may be significantly lower.

Our grid-based retrieval analysis does not identify a mass–metallicity trend in this hot Jupiter population. For a plausible correlation, we require a correlation coefficient greater than $|0.2|$ with a *p*-value greater than 0.05; however, from 10,000 simulations accounting for [M/H] uncertainty by sampling from each planet's [M/H] posterior probability distribution, the average Kendall rank correlation coefficient is $-0.16$ with a *p*-value of 0.40. This poor mass–metallicity correlation may indicate: (1) there are different formation and evolutionary pathways producing these hot Jupiters (e.g., Fortney et al. 2021), (2) elemental inventories vary significantly by star–planet system, and (3) more observations are required to tighten and improve our confidence in metallicity constraints. Thorngren et al. (2016) note a stronger mass–metallicity correlation when presenting metallicity relative to the host star than when presenting the metal fraction alone, although both correlate. In this manuscript, we do not normalize our metallicity constraints by their stellar hosts because uncertainties in both stellar metallicities and our atmosphere composition constraints limit how informative this stellar normalization can be.

C/O provides additional insight into planet formation (Öberg et al. 2011; Madhusudhan 2012). Referring to Figure 5, our grid-based retrievals do not constrain C/O in all cases, but for the planets with both upper and lower bounded constraints, roughly solar C/O is favored (darker shaded region in Figure 5). Four planets have upper bounds that include roughly solar composition or favor low C/O. Low C/O may suggest formation internal to the $H_2O$ ice line or O-rich solids dominating atmosphere accretion. Two planets favor sub- or supersolar C/O but not solar values, and one planet prefers supersolar only. If supersolar values are accurate, this may support gas-dominated envelope accretion beyond the $H_2O$ ice line followed by limited O-rich accretion and perhaps accretion of carbon grains, which have been proposed to exist at a "soot line" internal to the $H_2O$ ice line (e.g., Draine 2003). High C/O in hot Jupiter atmospheres has been suggested in prior literature. The finding of Welbanks et al. (2019) that $H_2O$/H abundances are much lower than $CH_4$/H for the same planets could be explained by high C/O (Madhusudhan et al. 2014a, 2014b). Other analyses of individual planets have also proposed high C/O compositions (e.g., WASP-18b: Sheppard et al. 2017, WASP-12b: Madhusudhan et al. 2011; Stevenson et al. 2014). However, other analyses and follow-up observations have not always supported these conclusions (e.g., WASP-18b: Coulombe et al. 2023, WASP-12b: this work).

The HST-only scenario results in more planets with unconstrained C/O and [M/H] or only upper/lower bounds. Broader spectral coverage in infrared wavelengths with JWST will continue to hugely improve composition constraints and strengthen our understanding of hot Jupiter formation pathways (e.g., Ahrer et al. 2023; Alderson et al. 2023; August et al. 2023; Bean et al. 2023; Coulombe et al. 2023; Feinstein et al. 2023; Kempton et al. 2023; Rustamkulov et al. 2023; Taylor et al. 2023).

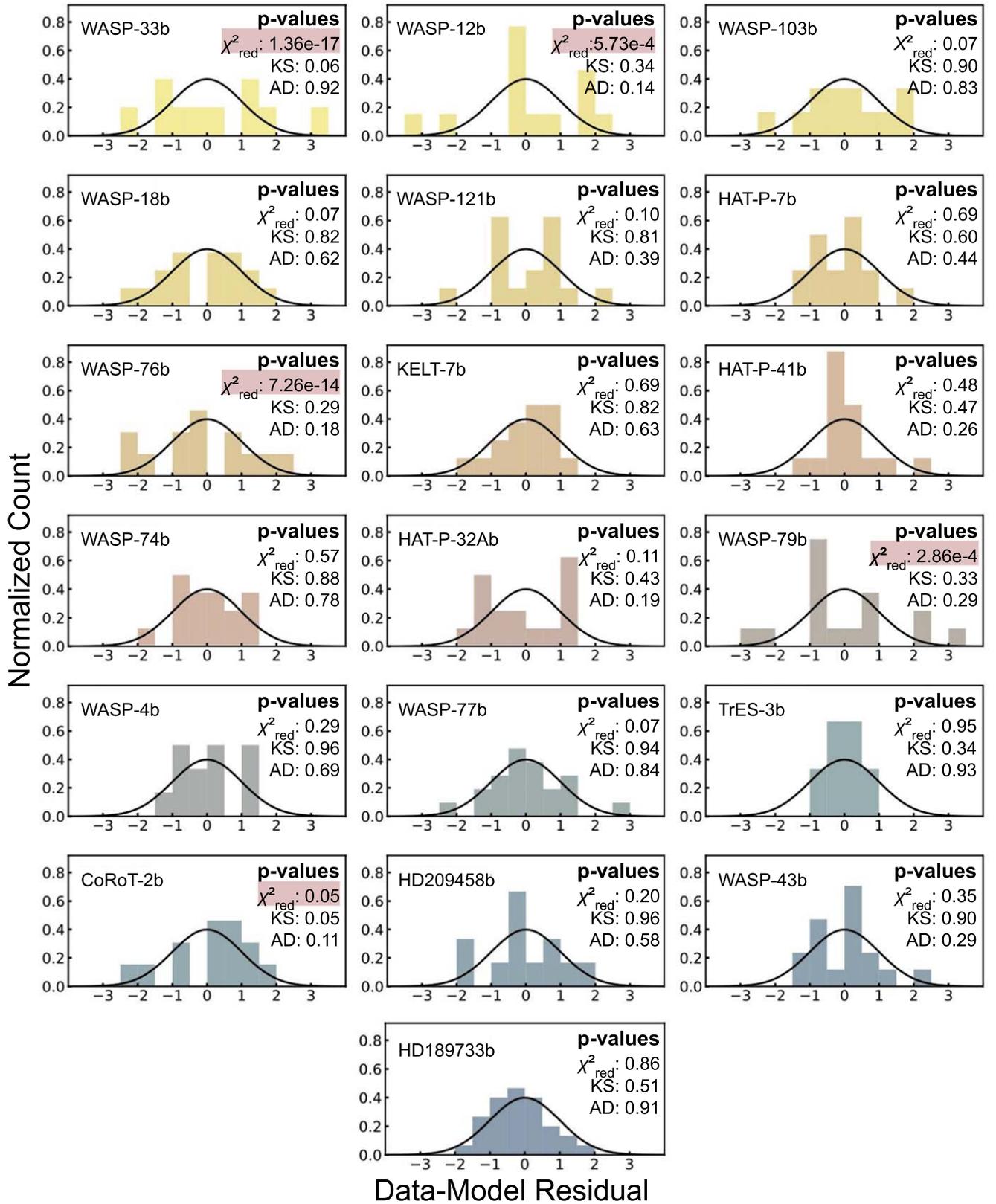

**Figure 3.** Histograms of data–model normalized residuals for the fiducial model scenario. Overplotted are normal distributions with mean = 0 and variance = 1. For ideal data–model agreement, random error results in scaled residuals drawn from a normal distribution about the model. Also reported in this figure are $p$-values calculated for the three statistical tests we perform: reduced-$\chi^2$ ($\chi^2_{\rm red}$), Kolmogorov–Smirnov (K-S), and Anderson–Darling (A-D). We take $p$-values > 0.05 to mean the model cannot be rejected by the observations. $P$-values ⩽ 0.05 are highlighted; however, we note that only the $\chi^2_{\rm red}$ test returns any $p$-values ⩽ 0.05 in the fiducial scenario. For all planets, the K-S and A-D tests cannot reject that these residuals are drawn from a normal distribution, so we do not reject the fiducial model assumptions.

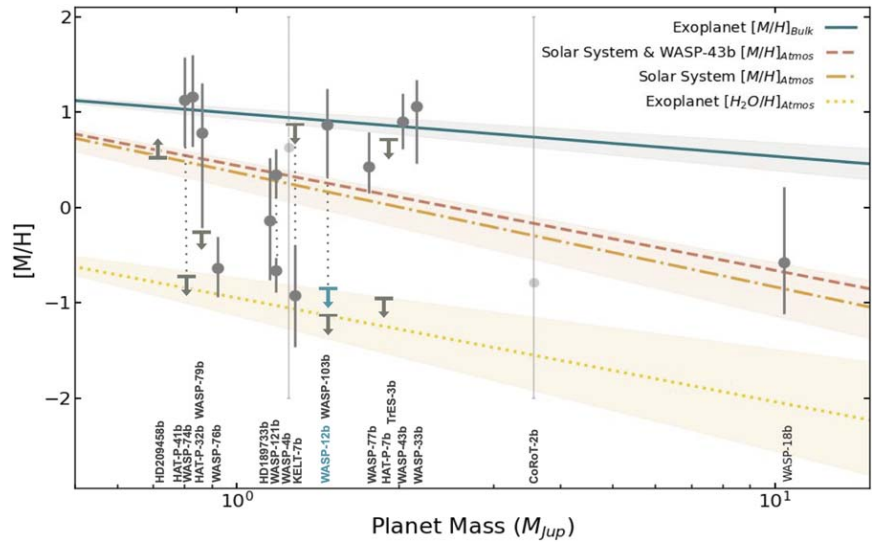

**Figure 4.** Atmosphere metallicity inferences relative to solar plotted alongside trend predictions from prior literature: exoplanet bulk metallicity relative to stellar (Thorngren et al. 2016), Solar System atmosphere metallicities derived from $CH_4$ abundances and WASP-43b atmosphere metallicity derived from $H_2O$ abundance relative to solar (Kreidberg et al. 2014), Solar System atmosphere metallicities (Guillot & Gautier 2015; Mordasini et al. 2016), and exoplanet atmosphere $H_2O$ abundance relative to stellar from transit spectra (Welbanks et al. 2019). Uncertainties on trends from prior literature are shaded (not available for the WASP-43b trend). Points are medians of the [M/H] posterior probability distributions from this work, error bars denote $1\sigma$ confidence regions, and dotted gray lines connect two possible solutions for a given planet (a bimodal posterior). When only an upper or lower constraint is found, an arrow is plotted at the $1\sigma$ confidence boundary. Faded points and error bars are shown for those planets with no metallicity constraint over the full grid range. WASP-12b is plotted in green to visually differentiate from WASP-103b. Our grid-based retrieval analysis does not identify a mass–metallicity trend in this hot Jupiter population. From 10,000 simulations accounting for uncertainty on [M/H] estimates by randomly drawing from each planet's [M/H] posterior probability distribution, the average Kendall rank correlation coefficient is $-0.16$, with an insignificant $p$-value of 0.40. With a correlation coefficient $<|0.2|$ and $p$-value $> 0.05$, we consider this no correlation.

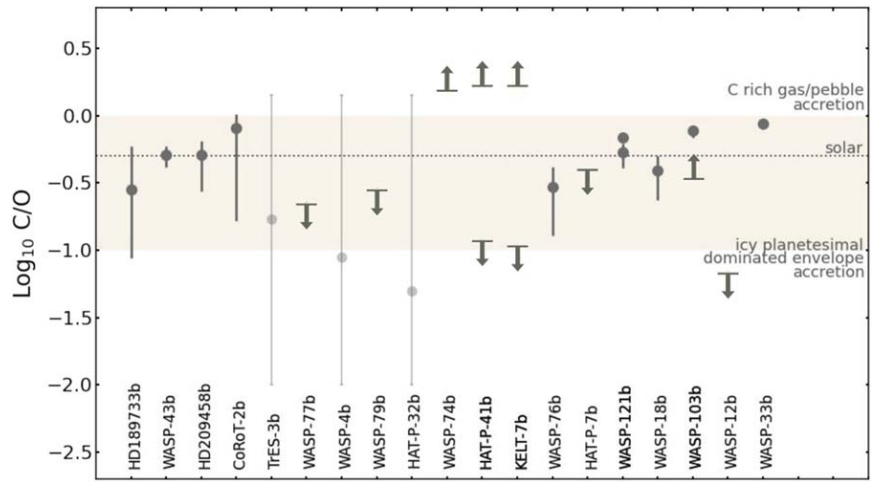

**Figure 5.** Carbon-to-oxygen ratio inferences ($\log_{10}C/O$) plotted alongside predicted ranges for different planet formation pathways. Points are medians of the $\log_{10}C/O$ posterior probability distributions, and error bars denote $1\sigma$ confidence regions. Faded points and error bars indicate planets with unconstrained C/O over the full grid range, and planets with two points have two possible solutions (a bimodal posterior). For planets with a partial constraint, an arrow is plotted at the $1\sigma$ boundary to denote an upper/lower limit. Planets are ordered by equilibrium temperature (not to scale). $\log_{10}C/O \gtrsim 0$ may indicate formation beyond the $H_2O$ ice line or further out, requiring carbon-rich gas and/or pebble accretion. Roughly solar C/O is consistent with formation beyond the $H_2O$ line with migration inward and some oxygen-rich accretion. Much lower C/O ($\log_{10}C/O \lesssim -1$) may indicate a high amount of oxygen-rich icy planetesimal accretion.

### 3.3. Planetary Climate

Constraints on longitudinal heat redistribution are presented in Figure 6 along with trends from GCMs with and without nightside clouds (Parmentier et al. 2021). We find that, for this data set, the heat redistribution parameter does not correlate with planet equilibrium temperature. We calculate Kendall rank correlations to assess the significance of a positive linear correlation, and from 10,000 simulations accounting for $f$ uncertainty by sampling from each planet's $f$ posterior probability distribution, the average Kendall correlation coefficient comes to 0.22 with an insignificant $p$-value of 0.23.

Interestingly, WASP-77b, WASP-4b, and WASP-79b have almost identical equilibrium temperatures and significantly different $f$ values; however, $f$ and the scale factor $A$ are degenerate. For instance, WASP-4b's high $f$ (see Figure 6) is accompanied with a low $A = 0.68^{+0.05}_{-0.04}$, but fixing $A = 1$ results in a lower $f = 1.86^{+0.11}_{-0.10}$. A deeper understanding of the mechanisms influencing heat transport in the atmospheres of hot Jupiters is necessary to accurately predict a planet's day-and nightside temperatures, in particular the presence and influence of aerosol species (e.g., Parmentier et al. 2016; Gao et al. 2021) and molecular dissociation (e.g., Tan & Komacek 2019; Mansfield et al. 2020; Roth et al. 2021).

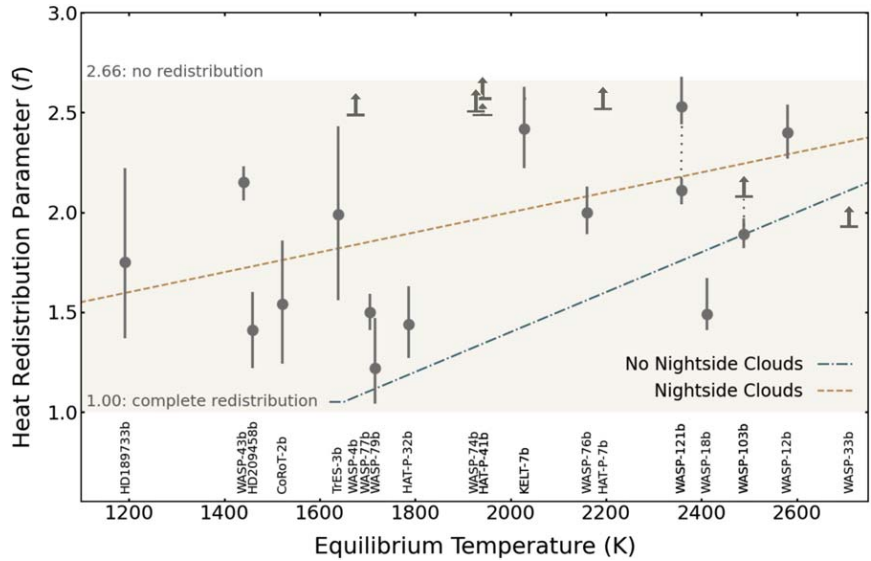

**Figure 6.** Heat redistribution ($f$) inferences plotted alongside trend predictions from GCM results in Parmentier et al. (2021). Points are medians of the $f$ posterior probability distributions, error bars denote $1\sigma$ confidence regions, and dotted gray lines connect two plausible solutions for a given planet (a bimodal posterior). When only an upper or lower constraint is found, an arrow is plotted at the $1\sigma$ confidence boundary. At the population level, we find an unlikely correlation between equilibrium temperature and $f$. From 10,000 simulations accounting for $f$ uncertainty by sampling from each planet's redistribution posterior probability distribution, the average Kendall rank correlation coefficient comes to 0.22 with an insignificant $p$-value of 0.24.

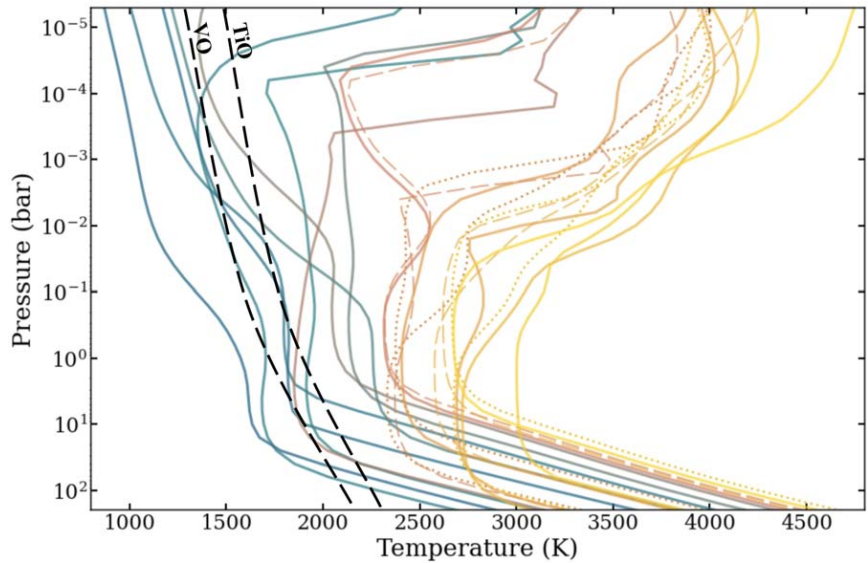

**Figure 7.** Modeled pressure–temperature profiles for each planet. Model parameters for each planet profile are consistent with posterior medians in the fiducial scenario. When two solutions exist, both $P$-$T$ profiles are shown in the same line style (dashed or dotted). Also shown are approximate TiO and VO condensation curves assuming solar elemental abundances. With increasing equilibrium temperatures (blue-to-yellow), temperature inversions develop at higher pressures.

In Figure 7, we plot pressure–temperature profiles for each planet. Temperature inversions are hypothesized to occur because of high-altitude absorbers (e.g., Fortney et al. 2008; Lothringer et al. 2018); modeled pressure–temperature profiles for this population are consistent with that trend. Condensation curves for TiO and VO are shown and fall roughly along the boundary between inverted and non-inverted atmospheres.

### 4. Discussion

Other recent studies have also aimed to summarize our understanding of hot Jupiters from the HST and Spitzer era. Here, we discuss how our results complement those presented in Goyal et al. (2021) and Changeat et al. (2022).

Goyal et al. (2021) estimates heat recirculation factors (Fortney & Marley 2007; Goyal et al. 2020), [M/H], and C/O for a population of 34 planets from Spitzer 3.6 and 4.5 $\mu$m emission observations. The authors use planet-specific, self-consistent model grids generated with the code ATMO. Goyal et al. (2021) consider nine of the same planets as our work, making our addition of HST WFC3 points complementary; however, we note that the Spitzer data presented in Goyal et al. (2021) are not fully consistent with the data we use, due to varied reduction methods. Additionally, the limits on C/O and [M/H] in the Goyal et al. (2021) grids are significantly narrower (limits noted in Figure 8). Parameter estimates from our work often include values outside of the Goyal et al. (2021) limits. Finally, Goyal et al. (2021) do not use MCMC/nested sampling and interpolation, instead computing $\chi^2$ across their grid. In agreement with our grid-based retrievals, Goyal et al. (2021) do not identify trends in C/O or [M/H], noting that Spitzer's precision is not fine enough to produce precise constraints on either parameter. Our analysis demonstrates that including Spitzer data narrows composition constraints compared to HST alone, so it is unsurprising that we also do not achieve precise constraints on all planets or identify obvious trends in composition.

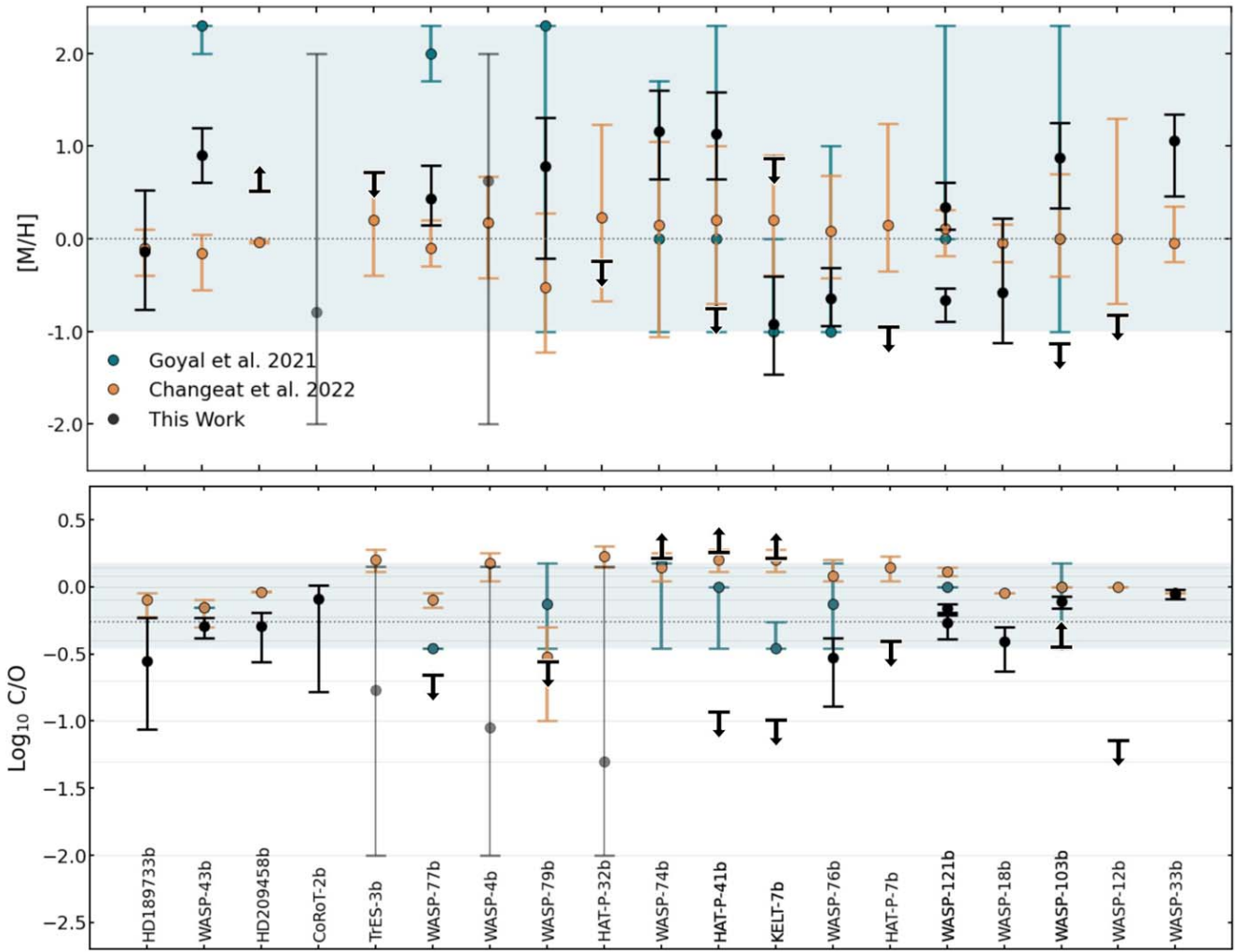

**Figure 8.** Parameter estimations from this work (black), Changeat et al. (2022) (orange), and Goyal et al. (2021) (blue). The Goyal et al. (2021) constraints are derived from grid models with narrower parameter ranges than this work, shown as shaded blue regions. Their parameter estimates are based on $\chi^2$ computed at each grid point, with error bars encompassing nearby grid points within $1\sigma$ $\chi^2$; therefore, points without error bars do not equate to infinite constraints but only indicate that no other grid points exist within $1\sigma$ confidence. The Changeat et al. (2022) constraints are derived from chemical equilibrium retrievals. Horizontal dotted lines convey solar composition. Top panel: [M/H]. Planet-specific grids presented in this manuscript have grid points in intervals of 0.2 dex. Bottom panel: $\log_{10}$C/O. Horizontal gray lines indicate grid intervals for our planet-specific grids. In the cases of WASP-74b, HAT-P-41b, and KELT-7b, grids were extended to $\log_{10}$C/O = 0.7.

Changeat et al. (2022) uniformly apply 1D retrieval tools for an analysis of 25 hot Jupiter emission spectra from HST and Spitzer. Eighteen of those planets overlap with our work; however, the data come from different reduction pipelines. The authors conducted free retrievals with the *Alfnor* and *TauREx3* codes (Changeat et al. 2020; Al-Refaie et al. 2021), as well as equilibrium chemistry retrievals using *GGChem* (Woitke et al. 2018). Changeat et al. (2022) find that free retrievals often produce a larger Bayesian evidence than chemical equilibrium, but in most cases, the Bayes factor comparison of the two model assumptions is nondecisive ($\Delta \ln(E) < 5$, the natural logarithm of the evidence, as defined in Changeat et al. 2022). This is consistent with our finding that chemical equilibrium cannot be ruled out.

Figure 8 compares parameter estimations from this work to equilibrium chemistry retrievals from Goyal et al. (2021) and Changeat et al. (2022). Agreements increase our statistical confidence in a given planet's parameter estimations from each study. However, significant differences are common and emphasize the subjective nature of applying different data–model inference methods to emission observations with HST and Spitzer. In general, the grid-based retrieval results from our five presented scenarios suggest that, with these observations alone, we cannot rule out the 1D radiative–convective–thermochemical equilibrium (1D-RCTE) assumption and solar composition atmospheres for hot Jupiter emission spectra.

### 4.1. Future Work

In this analysis, we use a Bayesian statistical framework for estimating atmosphere parameters; however, we adopt a frequentist statistical approach in our assessment of the quality of data–model fits at the population level. We take the HST and Spitzer data sets at face value, assuming no measurement uncertainties beyond random photon noise. We recognize that there are downsides to this assumption and our frequentist approach, mainly that, by adopting a global goodness-of-fit metric ($p$-values calculated from the $\chi^2_{\rm red}$, K-S, and A-D tests), we do not consider the reliance of data–model fits on individual data points (Welbanks et al. 2023). It is possible that data outliers or systematic uncertainties, such as instrument offsets and under/overestimated error bars, may be influencing our results unaccounted for. However, there is no statistical cause for assuming there are outliers in our considered data. From the K-S and A-D tests, $p$-values are generally greater than 0.05, meaning we cannot reject that residuals are drawn from a normal distribution and that there are no outliers in the data. Additionally, our primary conclusion that the 1D-RCTE assumption cannot be rejected for hot and ultra-hot Jupiters would not be altered by the consideration of additional measurement uncertainties. Future analyses that are able to reject model scenarios, such as with JWST and other space- and ground-based observatories, may conclude that outliers or systematic uncertainties are present, at which point methods such as the leave-one-out cross-validation presented in Welbanks et al. (2023) should be considered.

### 5. Conclusion

As we enter the era of JWST, our work summarizes what we learned from HST and Spitzer exoplanet emission spectra. We apply a uniform grid-based retrieval method to one of the most studied exoplanet populations, thus far, hot and ultra-hot Jupiters. We summarize key takeaways here:

1. Close data–model agreement is achievable with 1D, self-consistent, radiative–convective–thermochemical equilibrium grid models. This does not exclude complex chemistry and dynamics from existing, but suggests that specific claims may require additional observations.
2. Solar composition cannot be rejected for hot and ultra-hot Jupiter atmospheres. Future spectroscopic observations, in particular those longward of the HST WFC3 bandpass, are likely to result in more precise C/O and metallicity constraints.
3. From our analysis, a mass–metallicity Kendall rank correlation coefficient less than $|0.2|$ does not indicate a trend. We plot our constraints alongside hypothesized trends for context, but because of different analysis methods (i.e., stellar versus solar comparisons, M/H versus $H_2O$/H, and emission versus transmission observations), we do not support or reject the trends identified in prior work. Future observations may help to constrain M/H and reveal a trend. Alternatively, perhaps varied formation pathways do not facilitate a mass–metallicity relationship.
4. Heat redistribution estimates do not indicate a correlation between hotter equilibrium temperatures and increasingly restricted heat redistribution (Kendall rank correlation coefficient = 0.22, $p$-value = 0.24). Future analysis with additional data may improve heat redistribution constraints. Alternatively, climate variables such as cloud compositions, molecular dissociation, and winds may be varied enough to prevent a heat redistribution–temperature trend.

Results from this work show that the 1D-RCTE assumption is valid for hot and ultra-hot Jupiters when observed with HST WFC3 and Spitzer IRAC 3.6 and 4.5 $\mu$m. With our homogeneous modeling analysis of HST and Spitzer, we have built a foundation for our understanding of heat transport, atmospheric composition, and planet formation and migration from the pre-JWST data era. With JWST's added wavelength coverage, we anticipate increasingly precise constraints on composition and climate variables, illuminating further the drivers of planetary climate, chemistry, and planet formation pathways (e.g., Ahrer et al. 2023; Alderson et al. 2023; August et al. 2023; Bean et al. 2023; Bell et al. 2023; Feinstein et al. 2023; Kempton et al. 2023; Rustamkulov et al. 2023; Taylor et al. 2023). We also foresee observations of hot and ultra-hot Jupiters that cannot be well matched with the 1D-RCTE assumption, requiring the introduction of 3D and disequilibrium effects, as well as more rigorous statistical methods such as leave-one-out cross-validation. The results presented in this manuscript provide a starting point for improving our understanding of exoplanet atmospheres with future ground- and space-based telescopes.

### Acknowledgments

M.R.L. and L.S.W. acknowledge support from NASA/STScI award AR-16139, NASA Exoplanets Research Program grant 80NSSC19K0446, and the Nexus for Exoplanet System Science and NASA Astrobiology Institute Virtual Planetary Laboratory (No. 80NSSC18K0829). We also acknowledge Research Computing at Arizona State University for providing HPC and storage resources that have significantly contributed to the research results reported within this manuscript.

### Appendix

A table of grid model parameters (Table 2), modeled spectra zoomed in on the HST WFC3 wavelengths (Figure 9), fit statistics for all data-model scenarios (Table 3 and Figure 10), and parameter estimations for the fiducial model scenario (Table 4) are provided here. For a complete record of parameter estimations in each data-model scenario, see the online Zenodo repository (Wiser & Line 2024).

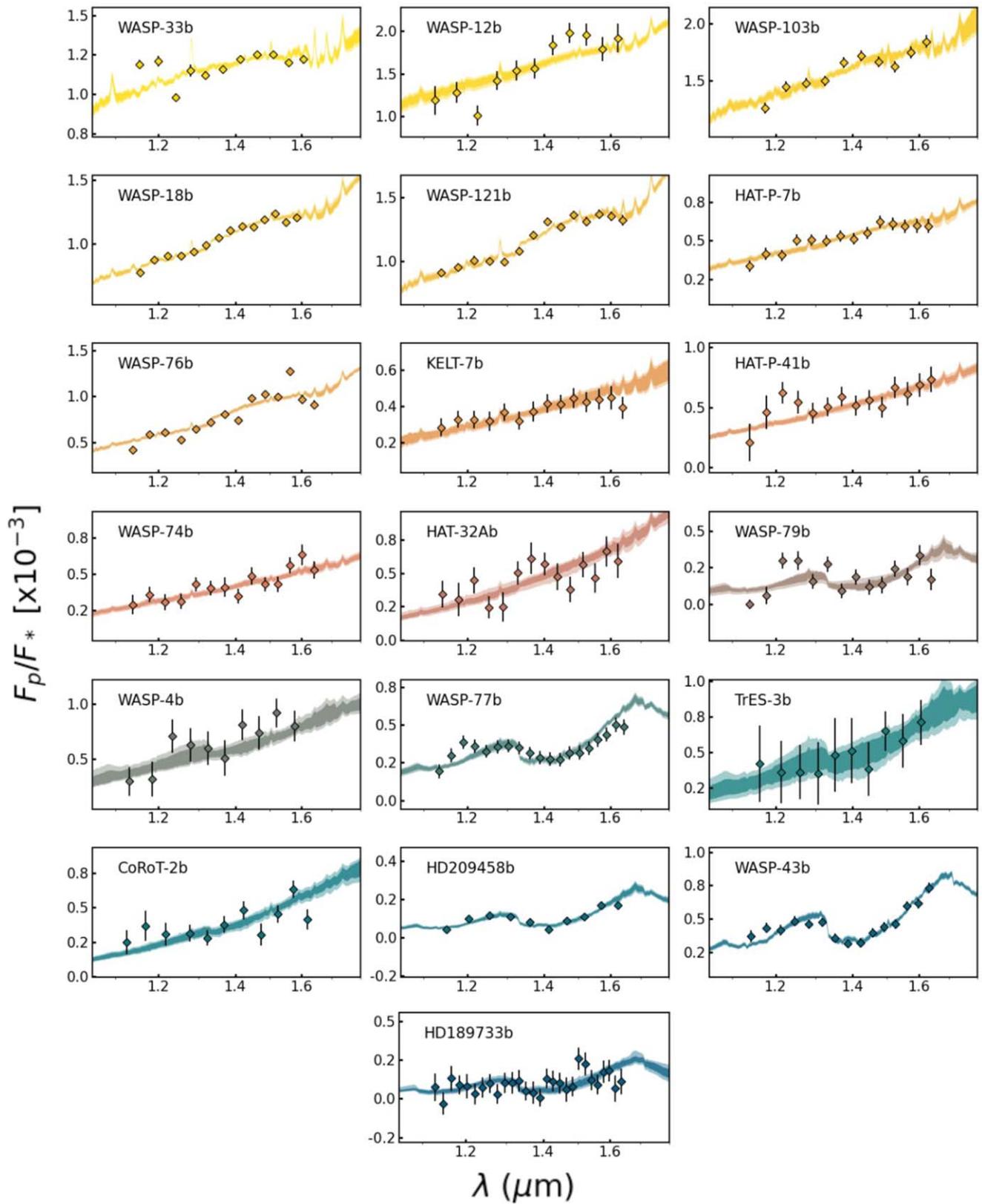

**Figure 9.** Modeled spectra for all 19 planets, zoomed in on the HST WFC3 wavelength range. The color bar correlates with the planet's equilibrium temperature, with higher temperatures toward the top of the figure. The 1σ and 2σ confidence regions calculated from Bayesian inference are shaded with darker and lighter shades, respectively. Corresponding corner plots for each planet can be found on the Zenodo repository.

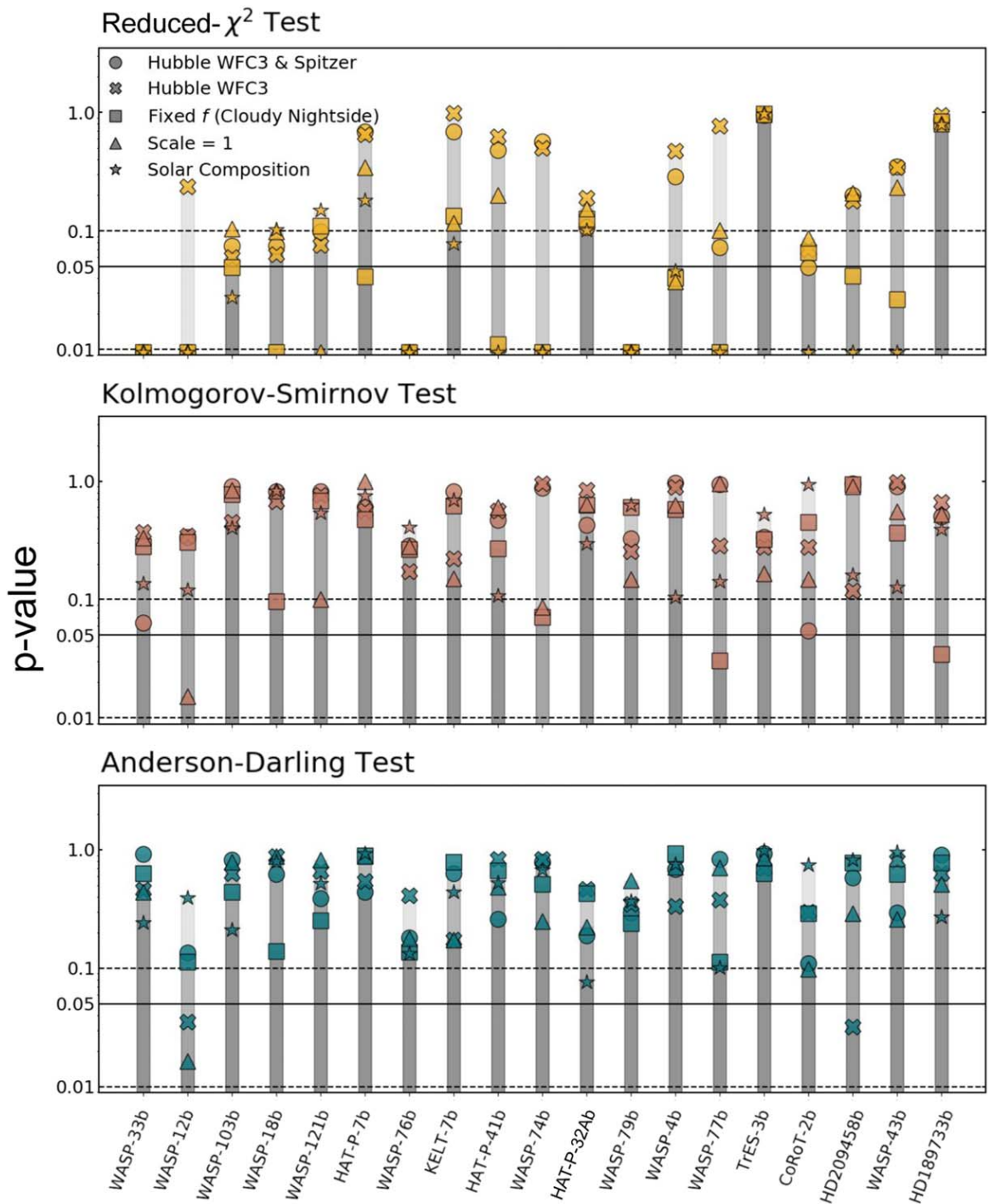

**Figure 10.** *P*-values calculated from three different statistical tests for the "goodness of fit" of modeled spectra to planet emission observations. In this study, we take *p*-values > 0.05 to mean the model assumptions cannot be ruled out as an explanation for the data. We label other common confidence thresholds, 0.1 and 0.01, with dashed lines. The $\chi^2_{\rm red}$ test results in the most model rejections, but only one data–model scenario is rejected by all three tests (WASP-12b, $A = 1$ scenario). Our primary conclusion is that the 1D-RCTE assumption cannot be ruled out for hot Jupiter atmospheres with these HST and Spitzer observations.

Table 2
Grid Properties

| Planet | Grid Ranges | | | | Planet Parameters | | | | Stellar Parameters | | | Sources |
|---|---|---|---|---|---|---|---|---|---|---|---|---|
| | [M/H] | C/O | $f$ | # Models | $T_{eq}$ (K) | $R_{Jup}$ | $M_{Jup}$ | $D$ (AU) | $T$ (K) | $R_{Sun}$ | log $g$ (c.g.s.) | |
| HD189733b | −2.0–2.4 | 0.01–1.4 | 0.8–2.6 | 2300 | 1191 | 1.151 | 1.15 | 0.03142 | 4969 | 0.784 | 4.51 | Krenn et al. (2023) |
| WASP-43b | −2.0–2.0 | 0.01–1.4 | 0.8–2.6 | 2100 | 1440 | 1.08 | 2.034 | 0.01526 | 4240 | 0.698 | 4.71 | Scandariato et al. (2022) |
| HD209458b | −2.0–2.2 | 0.01–1.4 | 0.8–2.8 | 2420 | 1459 | 1.38 | 0.714 | 0.04747 | 6117 | 1.162 | 4.368 | Southworth (2010) |
| CoRoT-2b | −2.0–2.0 | 0.01–1.6 | 0.8–2.8 | 2541 | 1521 | 1.46 | 3.57 | 0.02835 | 5598 | 0.901 | 4.527 | Southworth (2012) |
| TrES-3b | −2.0–2.0 | 0.01–1.4 | 0.6–2.8 | 2520 | 1638 | 1.31 | 1.899 | 0.02276 | 5650 | 0.8235 | 4.571 | Southworth (2011) |
| WASP-4b | −2.0–2.0 | 0.01–1.6 | 0.6–2.8 | 2772 | 1673 | 1.364 | 1.249 | 0.02318 | 5540 | 0.91 | 4.487 | Southworth (2012) |
| WASP-77b | −2.0–2.6 | 0.01–1.4 | 0.4–2.8 | 3120 | 1705 | 1.21 | 1.76 | 0.024 | 5605 | 0.955 | 4.33 | Maxted et al. (2013b) |
| WASP-79b | −2.0–2.0 | 0.01–1.4 | 0.4–2.8 | 2457 | 1716 | 1.53 | 0.86 | 0.0519 | 6600 | 1.51 | 4.226 | Brown et al. (2017) |
| HAT-P-32b | −2.0–2.0 | 0.01–1.6 | 0.8–2.8 | 2541 | 1786 | 1.789 | 0.86 | 0.0343 | 6269 | 1.219 | 4.335 | Nikolov et al. (2018) |
| | | | | | | | | | | | | Czesla et al. (2022) |
| WASP-74b | −2.0–2.0 | 0.01-5.0 | 0.8–2.8 | 2215 | 1926 | 1.404 | 0.826 | 0.03443 | 5986 | 1.536 | 4.141 | Mancini et al. (2019) |
| HAT-P-41b | −2.0–2.0 | 0.01-5.0 | 0.8–2.8 | 3234 | 1941 | 1.685 | 0.80 | 0.0426 | 6390 | 1.683 | 4.14 | Hartman et al. (2012) |
| KELT-7b | −2.0–2.0 | 0.01–5.0 | 1.0–2.8 | 2730 | 2028 | 1.496 | 1.28 | 0.0434 | 6699 | 1.712 | 4.15 | Tabernero et al. (2022) |
| WASP-76b | −2.0–2.0 | 0.01–1.4 | 0.8–2.8 | 2310 | 2160 | 1.83 | 0.92 | 0.033 | 6250 | 1.73 | 4.128 | West et al. (2016) |
| HAT-P-7b | −2.0–2.0 | 0.01–1.6 | 1.0–2.8 | 2310 | 2194 | 1.525 | 1.87 | 0.03805 | 6310 | 2.02 | 4.029 | Benomar et al. (2014) |
| WASP-121b | −2.0–2.0 | 0.01–1.4 | 0.8–2.8 | 2310 | 2358 | 1.865 | 1.183 | 0.02544 | 6586 | 1.44 | 4.47 | Borsa et al. (2021) |
| WASP-18b | −2.0–2.0 | 0.01–2.0 | 1.0–2.8 | 2730 | 2411 | 1.2159 | 10.38 | 0.016258 | 6435 | 1.249 | 4.35 | Coulombe et al. (2023) |
| WASP-103b | −2.0–2.0 | 0.01–1.6 | 0.6–2.8 | 2772 | 2489 | 1.646 | 1.47 | 0.01979 | 6110 | 1.413 | 4.219 | Delrez et al. (2018) |
| WASP-12b | −2.0–2.0 | 0.01–1.4 | 0.6–2.8 | 2520 | 2580 | 1.90 | 1.47 | 0.0234 | 6313 | 1.657 | 4.157 | Collins et al. (2017) |
| WASP-33b | −2.0–2.0 | 0.01–1.4 | 1.0–2.8 | 2100 | 2710 | 1.679 | 2.16 | 0.0259 | 7430 | 1.509 | 4.27 | Lehmann et al. (2015) |

**Notes.** Presented are grid ranges, resolution, and planet parameters used to generate atmosphere models. Planet and star parameters are sourced from TEPCat (Southworth 2011), and references cited by TEPCat are presented. Stellar spectra are interpolated from PHOENIX stellar models. In addition to the range of grid parameters, data–model fits allow for a scale factor ($A$) on $F_p/F_*$ with possible values of 0.5–1.5.

Table 3
Fit Statistics

| Planet Name | Scenario | $\chi^2_{\rm red}$ | p-value | K-S | p-value | A-D | p-value |
|---|---|---|---|---|---|---|---|
| WASP-33b | Fiducial | 11.15 | 1.36e-17 | 0.35 | 6.33e-02 | 0.18 | 9.15e-01 |
|  | WFC3 Only | 11.58 | 8.36e-15 | 0.26 | 3.73e-01 | 0.36 | 4.56e-01 |
|  | Fixed Solar | 13.27 | 9.75e-26 | 0.31 | 1.37e-01 | 0.47 | 2.43e-01 |
|  | Fixed $f$ | 10.01 | 5.28e-17 | 0.26 | 2.80e-01 | 0.28 | 6.30e-01 |
|  | $A = 1$ | 10.92 | 7.66e-19 | 0.25 | 3.32e-01 | 0.36 | 4.42e-01 |
| WASP-12b | Fiducial | 3.26 | 5.73e-04 | 0.25 | 3.35e-01 | 0.57 | 1.36e-01 |
|  | WFC3 Only | 1.32 | 2.38e-01 | 0.27 | 3.47e-01 | 0.82 | 3.52e-02 |
|  | Fixed Solar | 5.14 | 4.04e-08 | 0.31 | 1.21e-01 | 0.38 | 3.96e-01 |
|  | Fixed $f$ | 2.95 | 1.02e-03 | 0.26 | 3.06e-01 | 0.61 | 1.13e-01 |
|  | $A = 1$ | 6.53 | 3.54e-10 | 0.42 | 1.52e-02 | 0.95 | 1.64e-02 |
| WASP-103b | Fiducial | 1.78 | 7.48e-02 | 0.15 | 9.01e-01 | 0.22 | 8.28e-01 |
|  | WFC3 Only | 2.02 | 6.00e-02 | 0.26 | 4.48e-01 | 0.28 | 6.33e-01 |
|  | Fixed Solar | 2.02 | 2.74e-02 | 0.25 | 4.02e-01 | 0.50 | 2.12e-01 |
|  | Fixed $f$ | 1.88 | 4.94e-02 | 0.18 | 7.80e-01 | 0.36 | 4.42e-01 |
|  | $A = 1$ | 1.61 | 1.05e-01 | 0.17 | 8.44e-01 | 0.24 | 7.89e-01 |
| WASP-18b | Fiducial | 1.63 | 7.48e-02 | 0.15 | 8.17e-01 | 0.29 | 6.20e-01 |
|  | WFC3 Only | 1.75 | 6.34e-02 | 0.18 | 6.76e-01 | 0.20 | 8.78e-01 |
|  | Fixed Solar | 1.50 | 1.03e-01 | 0.15 | 8.29e-01 | 0.23 | 8.02e-01 |
|  | Fixed $f$ | 2.83 | 4.40e-04 | 0.30 | 9.61e-02 | 0.57 | 1.39e-01 |
|  | $A = 1$ | 1.53 | 9.92e-02 | 0.15 | 8.32e-01 | 0.20 | 8.83e-01 |
| WASP-121b | Fiducial | 1.55 | 9.88e-02 | 0.15 | 8.14e-01 | 0.39 | 3.88e-01 |
|  | WFC3 Only | 1.69 | 7.58e-02 | 0.17 | 7.62e-01 | 0.27 | 6.64e-01 |
|  | Fixed Solar | 1.39 | 1.50e-01 | 0.19 | 5.38e-01 | 0.33 | 5.20e-01 |
|  | Fixed $f$ | 1.49 | 1.11e-01 | 0.17 | 6.80e-01 | 0.47 | 2.52e-01 |
|  | $A = 1$ | 7.51 | 4.83e-15 | 0.29 | 1.01e-01 | 0.22 | 8.23e-01 |
| HAT-P-7b | Fiducial | 0.76 | 6.91e-01 | 0.18 | 6.00e-01 | 0.36 | 4.39e-01 |
|  | WFC3 Only | 0.78 | 6.52e-01 | 0.20 | 5.66e-01 | 0.31 | 5.44e-01 |
|  | Fixed Solar | 1.33 | 1.82e-01 | 0.16 | 7.51e-01 | 0.17 | 9.34e-01 |
|  | Fixed $f$ | 1.77 | 4.13e-02 | 0.20 | 4.76e-01 | 0.19 | 8.94e-01 |
|  | $A = 1$ | 1.11 | 3.45e-01 | 0.09 | 9.98e-01 | 0.20 | 8.85e-01 |
| WASP-76b | Fiducial | 7.88 | 7.26e-14 | 0.24 | 2.86e-01 | 0.53 | 1.80e-01 |
|  | WFC3 Only | 7.92 | 7.21e-13 | 0.28 | 1.72e-01 | 0.38 | 4.12e-01 |
|  | Fixed Solar | 7.20 | 2.82e-14 | 0.22 | 4.07e-01 | 0.58 | 1.33e-01 |
|  | Fixed $f$ | 7.25 | 1.83e-13 | 0.25 | 2.65e-01 | 0.57 | 1.37e-01 |
|  | $A = 1$ | 7.21 | 2.36e-13 | 0.24 | 2.81e-01 | 0.53 | 1.80e-01 |
| KELT-7b | Fiducial | 0.76 | 6.88e-01 | 0.15 | 8.23e-01 | 0.28 | 6.33e-01 |
|  | WFC3 Only | 0.25 | 9.91e-01 | 0.27 | 2.24e-01 | 0.53 | 1.73e-01 |
|  | Fixed Solar | 1.57 | 7.78e-02 | 0.17 | 6.92e-01 | 0.36 | 4.43e-01 |
|  | Fixed $f$ | 1.44 | 1.34e-01 | 0.18 | 6.20e-01 | 0.23 | 7.95e-01 |
|  | $A = 1$ | 1.48 | 1.16e-01 | 0.27 | 1.51e-01 | 0.53 | 1.74e-01 |
| HAT-P-41b | Fiducial | 0.96 | 4.81e-01 | 0.20 | 4.69e-01 | 0.46 | 2.61e-01 |
|  | WFC3 Only | 0.80 | 6.25e-01 | 0.20 | 5.60e-01 | 0.22 | 8.31e-01 |
|  | Fixed Solar | 2.13 | 8.21e-03 | 0.29 | 1.08e-01 | 0.32 | 5.29e-01 |
|  | Fixed $f$ | 2.11 | 1.10e-02 | 0.24 | 2.68e-01 | 0.27 | 6.68e-01 |
|  | $A = 1$ | 1.30 | 2.01e-01 | 0.18 | 6.03e-01 | 0.35 | 4.84e-01 |
| WASP-74b | Fiducial | 0.88 | 5.65e-01 | 0.14 | 8.76e-01 | 0.24 | 7.84e-01 |
|  | WFC3 Only | 0.93 | 5.02e-01 | 0.13 | 9.54e-01 | 0.22 | 8.33e-01 |
|  | Fixed Solar | 2.83 | 2.88e-04 | 0.41 | 5.33e-03 | 0.27 | 6.74e-01 |
|  | Fixed $f$ | 2.30 | 4.81e-03 | 0.31 | 7.07e-02 | 0.33 | 5.12e-01 |
|  | $A = 1$ | 3.04 | 1.69e-04 | 0.30 | 8.64e-02 | 0.47 | 2.51e-01 |
| HAT-32Ab | Fiducial | 1.52 | 1.08e-01 | 0.21 | 4.28e-01 | 0.52 | 1.90e-01 |
|  | WFC3 Only | 1.36 | 1.91e-01 | 0.15 | 8.45e-01 | 0.35 | 4.64e-01 |
|  | Fixed Solar | 1.50 | 1.01e-01 | 0.23 | 2.95e-01 | 0.68 | 7.64e-02 |
|  | Fixed $f$ | 1.45 | 1.27e-01 | 0.18 | 6.26e-01 | 0.37 | 4.27e-01 |
|  | $A = 1$ | 1.39 | 1.55e-01 | 0.18 | 6.41e-01 | 0.49 | 2.23e-01 |
| WASP-79b | Fiducial | 3.03 | 2.86e-04 | 0.23 | 3.26e-01 | 0.44 | 2.93e-01 |
|  | WFC3 Only | 3.63 | 7.58e-05 | 0.26 | 2.57e-01 | 0.41 | 3.48e-01 |

Table 3
(Continued)

| Planet Name | Scenario | $\chi^2_{\rm red}$ | $p$-value | K-S | $p$-value | A-D | $p$-value |
|---|---|---|---|---|---|---|---|
|  | Fixed Solar | 2.80 | 3.44e-04 | 0.18 | 6.27e-01 | 0.40 | 3.66e-01 |
|  | Fixed $f$ | 3.19 | 8.18e-05 | 0.18 | 6.05e-01 | 0.48 | 2.38e-01 |
|  | $A=1$ | 3.05 | 1.58e-04 | 0.27 | 1.48e-01 | 0.31 | 5.47e-01 |
| WASP-4b | Fiducial | 1.21 | 2.87e-01 | 0.13 | 9.62e-01 | 0.27 | 6.86e-01 |
|  | WFC3 Only | 0.93 | 4.74e-01 | 0.17 | 8.95e-01 | 0.42 | 3.33e-01 |
|  | Fixed Solar | 1.86 | 4.58e-02 | 0.34 | 1.04e-01 | 0.25 | 7.60e-01 |
|  | Fixed $f$ | 1.96 | 4.01e-02 | 0.21 | 5.82e-01 | 0.17 | 9.28e-01 |
|  | $A=1$ | 1.98 | 3.73e-02 | 0.20 | 6.32e-01 | 0.26 | 7.15e-01 |
| WASP-77b | Fiducial | 1.54 | 7.26e-02 | 0.11 | 9.41e-01 | 0.22 | 8.37e-01 |
|  | WFC3 Only | 0.71 | 7.73e-01 | 0.22 | 2.84e-01 | 0.39 | 3.80e-01 |
|  | Fixed Solar | 2.48 | 3.41e-04 | 0.24 | 1.43e-01 | 0.63 | 1.02e-01 |
|  | Fixed $f$ | 2.06 | 5.09e-03 | 0.31 | 3.06e-02 | 0.61 | 1.13e-01 |
|  | $A=1$ | 1.44 | 1.01e-01 | 0.11 | 9.50e-01 | 0.26 | 7.10e-01 |
| TrES-3b | Fiducial | 0.34 | 9.51e-01 | 0.26 | 3.35e-01 | 0.17 | 9.29e-01 |
|  | WFC3 Only | 0.23 | 9.65e-01 | 0.30 | 2.76e-01 | 0.27 | 6.92e-01 |
|  | Fixed Solar | 0.30 | 9.82e-01 | 0.22 | 5.22e-01 | 0.13 | 9.79e-01 |
|  | Fixed $f$ | 0.31 | 9.72e-01 | 0.26 | 3.23e-01 | 0.28 | 6.34e-01 |
|  | $A=1$ | 0.33 | 9.67e-01 | 0.31 | 1.67e-01 | 0.21 | 8.63e-01 |
| CoRoT-2b | Fiducial | 1.89 | 4.91e-02 | 0.36 | 5.48e-02 | 0.61 | 1.10e-01 |
|  | WFC3 Only | 1.87 | 6.95e-02 | 0.28 | 2.79e-01 | 0.44 | 2.97e-01 |
|  | Fixed Solar | 2.61 | 2.48e-03 | 0.14 | 9.35e-01 | 0.25 | 7.42e-01 |
|  | Fixed $f$ | 1.74 | 6.64e-02 | 0.23 | 4.49e-01 | 0.44 | 2.89e-01 |
|  | $A=1$ | 1.65 | 8.66e-02 | 0.30 | 1.49e-01 | 0.64 | 9.77e-02 |
| HD209458b | Fiducial | 1.38 | 2.00e-01 | 0.14 | 9.57e-01 | 0.30 | 5.83e-01 |
|  | WFC3 Only | 1.48 | 1.79e-01 | 0.36 | 1.20e-01 | 0.83 | 3.20e-02 |
|  | Fixed Solar | 2.57 | 4.10e-03 | 0.31 | 1.62e-01 | 0.22 | 8.24e-01 |
|  | Fixed $f$ | 1.94 | 4.16e-02 | 0.14 | 9.38e-01 | 0.24 | 7.78e-01 |
|  | $A=1$ | 1.34 | 2.08e-01 | 0.15 | 9.00e-01 | 0.44 | 2.90e-01 |
| WASP-43b | Fiducial | 1.10 | 3.51e-01 | 0.13 | 8.97e-01 | 0.44 | 2.93e-01 |
|  | WFC3 Only | 1.11 | 3.45e-01 | 0.11 | 9.78e-01 | 0.23 | 8.09e-01 |
|  | Fixed Solar | 2.55 | 8.25e-04 | 0.27 | 1.27e-01 | 0.15 | 9.62e-01 |
|  | Fixed $f$ | 1.85 | 2.64e-02 | 0.21 | 3.64e-01 | 0.29 | 6.26e-01 |
|  | $A=1$ | 1.24 | 2.34e-01 | 0.18 | 5.54e-01 | 0.46 | 2.61e-01 |
| HD189733b | Fiducial | 0.71 | 8.61e-01 | 0.15 | 5.06e-01 | 0.19 | 9.07e-01 |
|  | WFC3 Only | 0.57 | 9.54e-01 | 0.13 | 6.69e-01 | 0.28 | 6.29e-01 |
|  | Fixed Solar | 0.77 | 7.96e-01 | 0.16 | 3.94e-01 | 0.45 | 2.73e-01 |
|  | Fixed $f$ | 0.74 | 8.36e-01 | 0.25 | 3.48e-02 | 0.24 | 7.79e-01 |
|  | $A=1$ | 0.76 | 8.09e-01 | 0.14 | 5.23e-01 | 0.33 | 5.11e-01 |

**Note.** These statistics are plotted in Figure 10.

**Table 4**
Parameter Estimations—Fiducial Scenario

| Planet | $f$ | $f$ +err | $f$ −err | Limit | [M/H] | +err | −err | Limit | $\log_{10}(C/O)$ | +err | −err | Limit | $A$ | +err | −err |
|---|---|---|---|---|---|---|---|---|---|---|---|---|---|---|---|
| HD189733b | 1.75 | 0.47 | 0.38 | ⋯ | −0.14 | 0.66 | 0.62 | ⋯ | −0.55 | 0.32 | 0.51 | ⋯ | 0.93 | 0.26 | 0.17 |
| WASP-43b | 2.15 | 0.08 | 0.09 | ⋯ | 0.90 | 0.30 | 0.29 | ⋯ | −0.29 | 0.06 | 0.09 | ⋯ | 0.89 | 0.04 | 0.03 |
| HD209458b | 1.41 | 0.19 | 0.19 | ⋯ | 1.04 | 0.69 | 0.53 | LL | −0.29 | 0.10 | 0.27 | ⋯ | 0.93 | 0.19 | 0.12 |
| CoRoT-2b | 1.54 | 0.32 | 0.30 | ⋯ | −0.79 | 2.22 | 1.02 | ⋯ | −0.09 | 0.10 | 0.69 | ⋯ | 1.21 | 0.17 | 0.17 |
| TrES-3b | 1.99 | 0.44 | 0.43 | ⋯ | −0.60 | 1.31 | 0.96 | UL | −0.77 | 0.60 | 0.89 | UC | 0.80 | 0.17 | 0.12 |
| WASP-77b | 1.50 | 0.09 | 0.09 | ⋯ | 0.43 | 0.36 | 0.28 | ⋯ | −1.19 | 0.54 | 0.53 | UL | 0.98 | 0.05 | 0.05 |
| WASP-4b | 2.66 | 0.10 | 0.17 | LL | 0.63 | 0.81 | 1.92 | UC | −1.05 | 0.72 | 0.64 | UC | 0.68 | 0.05 | 0.04 |
| WASP-79b | 1.22 | 0.25 | 0.18 | ⋯ | 0.78 | 0.53 | 0.99 | ⋯ | −1.29 | 0.74 | 0.48 | UL | 1.28 | 0.19 | 0.20 |
| HAT-P-32b | 1.44 | 0.19 | 0.17 | ⋯ | −1.03 | 0.78 | 0.64 | UL | −1.30 | 1.14 | 0.48 | UC | 1.11 | 0.09 | 0.09 |
| WASP-74b | 2.66 | 0.10 | 0.15 | LL | 1.16 | 0.44 | 0.52 | ⋯ | 0.45 | 0.17 | 0.25 | LL | 0.74 | 0.04 | 0.03 |
| HAT-P-41b | 2.66 | 0.10 | 0.16 | LL | 1.13 | 0.45 | 0.49 | ⋯ | 0.45 | 0.17 | 0.20 | LL | 0.87 | 0.05 | 0.04 |
|  | 2.71 | 0.06 | 0.14 | LL | −1.55 | 0.82 | 0.33 | UL | −1.45 | 0.52 | 0.38 | UL | 0.88 | 0.05 | 0.05 |
| KELT-7b | 2.42 | 0.21 | 0.20 | ⋯ | −0.10 | 0.98 | 1.34 | UL | −1.59 | 0.60 | 0.30 | UL | 0.88 | 0.05 | 0.05 |
|  | 2.69 | 0.08 | 0.12 | LL | −0.92 | 0.52 | 0.54 | ⋯ | 0.44 | 0.18 | 0.21 | LL | 0.79 | 0.03 | 0.02 |
| WASP-76b | 2.00 | 0.13 | 0.11 | ⋯ | −0.64 | 0.33 | 0.30 | ⋯ | −0.53 | 0.15 | 0.36 | ⋯ | 1.01 | 0.04 | 0.03 |
| HAT-P-7b | 2.66 | 0.09 | 0.14 | LL | −1.44 | 0.49 | 0.37 | UL | −1.05 | 0.64 | 0.64 | UL | 0.89 | 0.04 | 0.03 |
| WASP-121b | 2.53 | 0.15 | 0.09 | ⋯ | 0.34 | 0.27 | 0.24 | ⋯ | −0.27 | 0.06 | 0.12 | ⋯ | 0.69 | 0.02 | 0.02 |
|  | 2.11 | 0.06 | 0.07 | ⋯ | −0.66 | 0.13 | 0.23 | ⋯ | −0.16 | 0.03 | 0.03 | ⋯ | 0.74 | 0.02 | 0.03 |
| WASP-18b | 1.49 | 0.18 | 0.08 | ⋯ | −0.58 | 0.80 | 0.54 | ⋯ | −0.41 | 0.11 | 0.22 | ⋯ | 0.96 | 0.04 | 0.04 |
| WASP-103b | 1.89 | 0.08 | 0.07 | ⋯ | −1.59 | 0.45 | 0.26 | UL | −0.11 | 0.04 | 0.05 | ⋯ | 1.01 | 0.04 | 0.04 |
|  | 2.42 | 0.27 | 0.34 | LL | 0.87 | 0.38 | 0.54 | ⋯ | −0.24 | 0.34 | 0.22 | LL | 0.92 | 0.04 | 0.03 |
| WASP-12b | 2.40 | 0.14 | 0.13 | ⋯ | −1.36 | 0.50 | 0.40 | UL | −1.60 | 0.43 | 0.27 | UL | 0.78 | 0.03 | 0.03 |
| WASP-33b | 2.41 | 0.26 | 0.48 | LL | 1.06 | 0.28 | 0.60 | ⋯ | −0.06 | 0.04 | 0.03 | ⋯ | 0.77 | 0.08 | 0.04 |

**Notes.** Unconstrained parameters are labeled with "UC," and parameters with upper or lower limits are labeled "UL" and "LL" respectively. Parameter estimations in other retrieval scenarios can be found in a complete table on Zenodo.

### ORCID iDs

Lindsey S. Wiser https://orcid.org/0000-0002-3295-1279
Michael R. Line https://orcid.org/0000-0002-2338-476X
Luis Welbanks https://orcid.org/0000-0003-0156-4564
Megan Mansfield https://orcid.org/0000-0003-4241-7413
Vivien Parmentier https://orcid.org/0000-0001-9521-6258
Jacob L. Bean https://orcid.org/0000-0003-4733-6532
Jonathan J. Fortney https://orcid.org/0000-0002-9843-4354